\documentclass[aps,prb,floatfix,twocolumn,superscriptaddress,longbibliography,10pt]{revtex4-2}
\usepackage{amsmath,amssymb,graphicx,amssymb}
\usepackage[colorlinks=true,citecolor=blue,linkcolor=red]{hyperref}
\usepackage{bm}
\usepackage{hyperref}
\usepackage{subfigure}
\usepackage{natbib}
\usepackage[dvipsnames]{xcolor}

\newcommand{\ket}[1]{\left| #1 \right\rangle}
\newcommand{\bra}[1]{\left\langle #1 \right|}

\newcommand{\expn}[1]{{\rm e}^{#1}}
\newcommand{\bv}[1]{\mathbf{#1}}
\newcommand{\dg}{{^{\dagger}}}

\newcommand{\eg}{\textit{e.g.,}~}

\newcommand{\nn}{\nonumber}

\graphicspath{ {./Figures/} }

\begin{document}
\title{Benchmarking low-power flopping-mode spin qubit fidelities in Si/SiGe devices with alloy disorder}

\author{Steve Young}
\email{syoung1@sandia.gov}
\affiliation{Center for Computing Research, Sandia National Laboratories, Albuquerque, New Mexico 87185, USA}

\author{Mitchell Brickson}
\affiliation{Center for Computing Research, Sandia National Laboratories, Albuquerque, New Mexico 87185, USA}

\author{Jason R. Petta}
\affiliation{Department of Physics and Astronomy, University of California--Los Angeles, Los Angeles, California 90095, USA}
\affiliation{Center for Quantum Science and Engineering, University of California--Los Angeles, Los Angeles,
California 90095, USA}

\author{N. Tobias Jacobson}
\email{ntjacob@sandia.gov}
\affiliation{Center for Computing Research, Sandia National Laboratories, Albuquerque, New Mexico 87185, USA}

\date{\today}
\begin{abstract}
In the ``flopping-mode" regime of electron spin resonance, a single electron confined in a double quantum dot is electrically driven in the presence of a magnetic field gradient. The increased dipole moment of the charge in the flopping mode significantly reduces the amount of power required to drive spin rotations. However, the susceptibility of flopping-mode spin qubits to charge noise, and consequently their overall performance, has not been examined in detail. In this work, we simulate single-qubit gate fidelities of electrically driven spin rotations in an ensemble of devices configured to operate in both the single-dot and flopping-mode regimes. Our model accounts for the valley physics of conduction band electrons in silicon and realistic alloy disorder in the SiGe barrier layers, allowing us to investigate device-to-device variability. We include charge and magnetic noise, as well as spin relaxation processes arising from charge noise and electron-phonon coupling. We find that the two operating modes exhibit significantly different susceptibilities to the various noise sources, with valley splitting and spin relaxation times also playing a role in their relative performance. For realistic noise strengths, we find that single-dot gate fidelities are limited by magnetic noise while flopping-mode fidelities are primarily limited by charge noise and spin relaxation. For sufficiently long spin relaxation times, flopping-mode spin operation is feasible with orders-of-magnitude lower drive power and gate fidelities that are on par with conventional single-dot electric dipole spin resonance.
\end{abstract}

\maketitle
\section{Introduction}
The performance of quantum dot-based spin qubits in silicon has steadily improved over the past decade \cite{burkard2023semiconductor}. Advances in device fabrication and heterostructure growth have increased valley splittings \cite{McJunkin2022,PaqueletWuetz2022,DegliEsposti2024}. Isotopic purification has reduced magnetic noise originating from the contact hyperfine interaction with lattice nuclei, resulting in higher gate fidelities \cite{Struck2020,Mills2022,Weinstein2023}. Recent progress demonstrating the operation of multiple spin qubits suggests that electron spins in silicon may indeed be a viable platform for future large-scale quantum computing \cite{Takeda2022,Philips2022,Neyens2024}.

As the number of qubits in a device increases, the power required to implement gate operations may limit overall performance.
For single-spin ``Loss-DiVincenzo" qubits, single-qubit rotations are driven magnetically, by driving a substantial current through a nearby stripline \cite{Koppens2006,Pla2012,veldhorst2015two,Hile2018} or by electrically driving the spin in the presence of a strong magnetic field gradient \cite{Tokura2006,Pioro-Ladriere2008} in a process now commonly referred to as electric dipole spin resonance (EDSR) \cite{burkard2023semiconductor}. Both of these approaches have yielded single-qubit gate fidelities exceeding $F$~$>$~99.9\% \cite{Muhonen2015,Yang2019,Mills2022,Noiri2022,Lawrie2023}. 
However, some puzzling non-linearities have been observed \cite{Yoneda2018,Undseth2023}.
It is still an open question if these single-spin driving approaches can be scaled to the 10$^6$ -- 10$^8$ physical qubits that will be required for fault-tolerant operation \cite{dalzell2023quantum}.
\begin{figure*}
		\centering
		\includegraphics[width=0.95\textwidth]{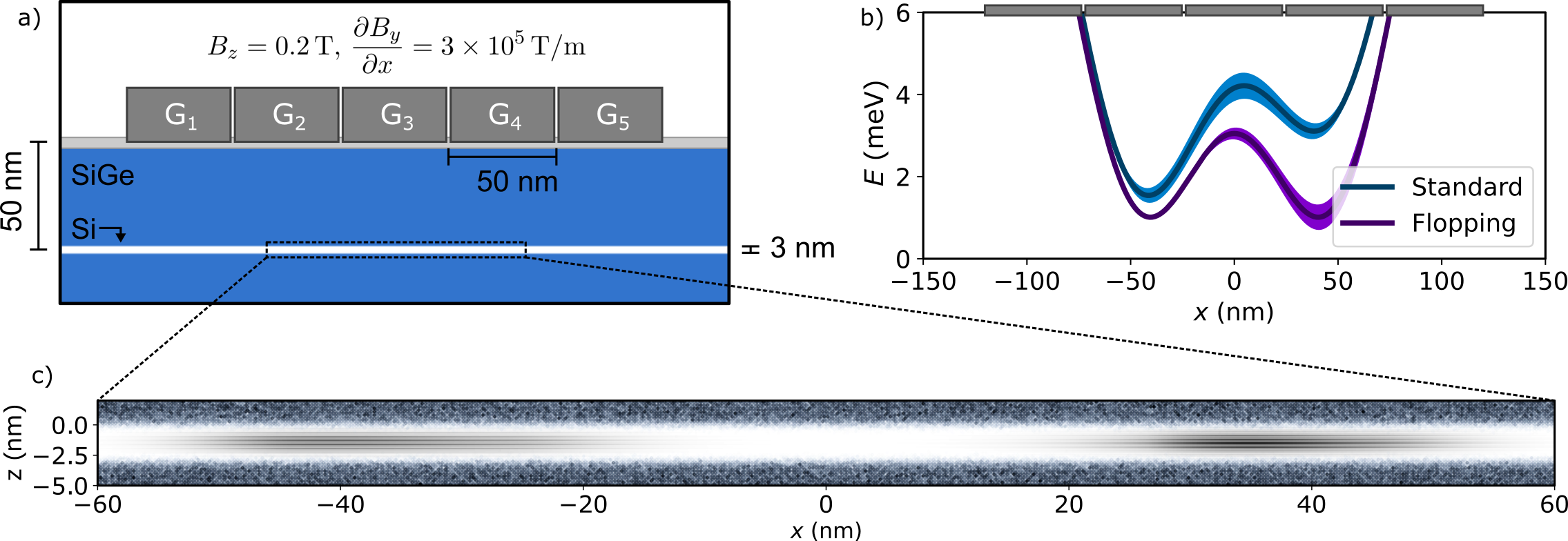}
		\caption{\textbf{Device layout.} (a) Cross-section of the device stack. The heterostructure consists of a 3 nm thick Si quantum well that is buried beneath a 50 nm thick Si$_{0.7}$Ge$_{0.3}$ spacer layer. Five gate electrodes, labeled G$_1$ -- G$_5$, generate the confinement potential in the plane of the quantum well.  (b) Electronic confinement potential $V(x)$ in SD and FM EDSR regimes. While operating in the SD EDSR mode, a single electron is isolated beneath either G$_2$ or G$_4$, whereas in the FM EDSR regime the electron is delocalized across a DQD formed beneath gates G$_2$ and G$_4$, with gate G$_3$ setting the interdot barrier height. In SD EDSR (FM EDSR) gate G$_3$ (G$_4$) is used to apply the microwave drive field, with the potential variation due to the ac drive illustrated by the shaded regions. (c) Magnified view of the quantum well structure, with white representing Si and blue representing Ge; darker shades of blue indicate higher Ge concentrations. The ground state FM charge density $\vert \psi(x,y=0,z) \vert^{2}$ is shown in black. } \label{fig:schematic}
	\end{figure*}
Practical limitations on the power required to drive fast single-spin rotations motivate the consideration of alternative lower-power modes of operation. One viable approach is to drive single-spin rotations in the ``flopping-mode'' regime of electric dipole spin resonance (FM EDSR), where a single electron spin is delocalized in a double quantum dot (DQD) \cite{Benito2019b,Croot2020}. The larger electronic dipole moment of the electron in the DQD allows the electron to experience a larger oscillating transverse magnetic field at much lower drive powers. Indeed, theoretical and experimental work has provided strong evidence that the flopping-mode allows for significant reduction in drive power ($\sim$10$^3$ less) in magnetic field gradients \cite{Benito2019b,Croot2020} as well as intrinsic spin-orbit fields in materials such as Ge \cite{Mutter2021}.
On the other hand, the enhanced charge dipole moment in FM EDSR may increase the susceptibility of the spin qubit to charge noise. Therefore, determining the overall fidelity of single-qubit gates driven with FM EDSR is of high importance. 

In this manuscript, we use numerical simulations to characterize the fidelities of single-qubit gates driven in both the conventional single-dot electric dipole spin resonance (SD EDSR) and low power FM EDSR regimes. To account for realistic alloy disorder effects that can reduce valley splittings, we evaluate the performance of an ensemble of devices operated in both regimes. We investigate potential issues that may arise due to small valley splittings, as well as the sensitivity of each operating mode to magnetic and charge noise. We furthermore examine the impact of spin relaxation ($T_1$) on gate fidelities. Our results show that for sufficiently long spin relaxation times, FM EDSR is feasible with orders-of-magnitude lower drive power and gate fidelities that are on par with conventional SD EDSR.

\section{Model and Methods}
\subsection{Device}
	
	\begin{table}
		\begin{centering}
			\begin{tabular}{|c||c|c|c|c|c|}
                \hline
                \multicolumn{6}{|c|}{\textbf{SD EDSR}}\\
                \hline
				Dot & $V_1$ (V) & $V_2$ (V) & $V_3$ (V) & $V_4$ (V) & $V_5$ (V)   \\
				\hline
                Left & 0.0 & 0.120 & 0.025 & 0.105 & 0.0\\
                Right & 0.0 & 0.105 & 0.025 & 0.120 & 0.0\\
				\hline
                \hline
                \multicolumn{6}{|c|}{\textbf{FM EDSR}}\\
                \hline
				Device & $V_1$ (V) & $V_2$ (V) & $V_3$ (V), $n=0...3$ & $V_4$ (V) & $V_5$ (V)   \\
				\hline
                1 & 0.0 & 0.120 & 0.0224+0.003$n$ & 0.117 & 0.0\\
                2 & 0.0 & 0.120 & 0.0222+0.003$n$ & 0.113 & 0.0\\
                3 & 0.0 & 0.120 & 0.0208+0.003$n$ & 0.119 & 0.0\\
                4 & 0.0 & 0.120 & 0.0224+0.003$n$ & 0.121 & 0.0\\
                5 & 0.0 & 0.120 & 0.0222+0.003$n$ & 0.121 & 0.0\\
				\hline
			\end{tabular}
			\caption{\textbf{Baseline idle gate voltages for all devices and operating modes.}  The top panel gives the voltages for dots under $G_2$ (left) and $G_4$ (right) for all devices. The bottom panel gives the voltages for FM EDSR for each device, which differed due to the need to tune to the zero-bias point. For each FM device four $V_3$ settings in 3 mV increments were used, resulting in different tunnel couplings.   }\label{tab:voltages}
		\end{centering}
	\end{table}

We simulate the SD and FM EDSR operating modes on devices with the structure depicted in Fig.~\ref{fig:schematic}. The device stack consists of five 50 nm-wide gate electrodes that are separated from a 3 nm wide Si quantum well by a 50~nm thick upper Si$_{0.7}$Ge$_{0.3}$ spacer layer. The SiGe layers are characterized by an explicit distribution of Ge defects within otherwise bulk Si~\cite{Pena2024}, resulting in a unique alloy disorder realization for each device.  We assume that the Si/SiGe interfaces are perfectly flat, with an intermixing length $4\tau = 1 \ \mathrm{nm}$ in the growth direction that is consistent with the characterization of Si/SiGe heterostructures in the literature \cite{Dyck2017,PaqueletWuetz2022,Pena2024}. The magnetic field is simulated by imposing a vertically oriented Zeeman field $B_z$~=~0.2~T and a horizontally oriented magnetic field gradient $\partial B_y/\partial x=3\times10^5$T/m across the device. 
Within our model, our results would be unchanged if $B_z$ were instead applied along the inter-dot axis, as may be typical in experiment \cite{Croot2020}.
    
\subsection{Microscopic Model}
	The electronic structure for each device and operating mode is determined using an envelope function formalism that accounts for alloy disorder in the SiGe barrier layer \cite{Pena2024}. Our approach entails using an envelope description of the electronic wavefunction on a hexahedral mesh making use of a discontinuous Galerkin discretization in terms of Cartesian products of Legendre polynomials \cite{HesthavenWarburtonBook}. To capture valley physics, we include previously calculated bulk Si Bloch functions \cite{Gamble2015}. In Appendix \ref{appendix:Laconic}, we give more details about how we perform electronic structure calculations using our code, \emph{Laconic}.
	
	The lowest four states at idle are collected and rotated into a localized basis using the Boys localization criteria \cite{Boys1960}, yielding four valley-orbital basis states.  The Hamiltonian elements are projected into this basis and the spin degree of freedom is added, resulting in an effective Hamiltonian near idle
	\begin{flalign}
		\hat{H}(\Delta\bv{V},\Delta\bv{B_z})&=\hat{H}^0+\sum_i\hat{H}_{{\rm G}_i}\Delta V_i+\sum_j\hat{H}_{M_{z,j}}\Delta B_{z,j}\nn\\
		\hat{H}^0&=\sum_i\hat{H}_{{\rm G}_i}V_i+B_z\sum_j\hat{H}_{M_{z,j}}+B_x\hat{H}_{M_x}\label{eq:ham}
	\end{flalign}
	where $\Delta \bv{V}$ and $\Delta \bv{B}_{z}$ represent the deviations of device parameters from idle values, $\hat{H}_{M_{x}}=\hat{x}\hat{S}_y$ is the coupling to the magnetic field gradient, and $\hat{H}_{M_{z,j}}$ and $\hat{H}_{V_i}$ couple the quantum dot(s) to the Zeeman fields $B_{z,j}$ and electrode voltages $V_i$, respectively.  In the SD EDSR case there is a single $\hat{H}_{M_{z,1}}=\gamma_{e} \hat{S}_z$, while in the FM EDSR case there is a Zeeman operator for the states localized in the left and right sites of the DQD,  $\hat{H}_{M_{z,L/R}}=\gamma_{e} \hat{P}_{L/R}\hat{S}_z$. Here $\hat{P}_{L/R}$ projects onto the left/right dot states and $\gamma_{e}~\approx~116~\mu$eV/T.
	
	\subsection{Dynamics and Noise}
    \label{sec:noise}
	We include multiple noise sources in our simulations, partitioning them into low- and high-frequency contributions, roughly corresponding to energies that are much lower than or near to, respectively, the qubit transition energy set by $B_z$ of $\sim 23~\mu$eV.  At low frequencies (1 Hz to 100 MHz) we incorporate both hyperfine and charge noise as $1/f$ processes modeled using ensembles of Ornstein-Uhlenbeck processes $\eta(t)=v\sum_{k=0}^7\int^t\expn{-\lambda_k\tau}dW_\tau$ where $\lambda_k=10^{-k}$~Hz. In the case of hyperfine noise, the noise processes contribute fluctuations to the Zeeman field(s), which are otherwise static, so that $\Delta B_{z,j}(t)=b_z\eta_j(t)$ where we take $b_z \approx 8.6 \times 10^{-8} \ {\rm{T/\sqrt{Hz}}}$.  In SD EDSR mode, a single process perturbs all states ($j=1$), while in FM EDSR mode two independent processes are applied ($j=L,R$) \cite{Hung2013}. Charge noise is approximated as ``gate referred'' and simulated by associating each electrode with an independent noise process that contributes fluctuations to $\Delta V_i(t)$.  To distinguish between noise and control we let $\Delta V_i(t)\rightarrow \Delta V_i(t)+ v\eta_i(t)$ where $\Delta V_i(t)$ is now the control and $v\eta_i(t)$ with $v=10^{-5}V/\sqrt{{\rm Hz}}$ are the noise contributions.
	
	These noise channels are incorporated into our dynamics by applying a ``dynamic'' quasi-static approximation.  We discretize the noise processes as fluctuating over intervals of a selected duration, here chosen to be 50~ns, and compute dynamics for this interval given a stationary Hamiltonian determined by the interval's associated noise values.  
	
	We include high-frequency ($T_1$) noise originating from both charge fluctuations and electron-phonon coupling, approximating both as Markovian and incorporated into the dynamics simulation as Lindblad processes.  High-frequency charge noise is also modeled as gate referred, with a Linbladian term for each $H_{G_i}$. We describe how we compute the phonon contribution to spin $T_1$ processes in Appendix \ref{appendix:Phonons} and the charge noise contribution in Appendix \ref{appendix:ChargeNoiseT1}.

	\subsection{Electronic Structure}
		\begin{figure}
		\centering
		\includegraphics[width=0.5\textwidth]{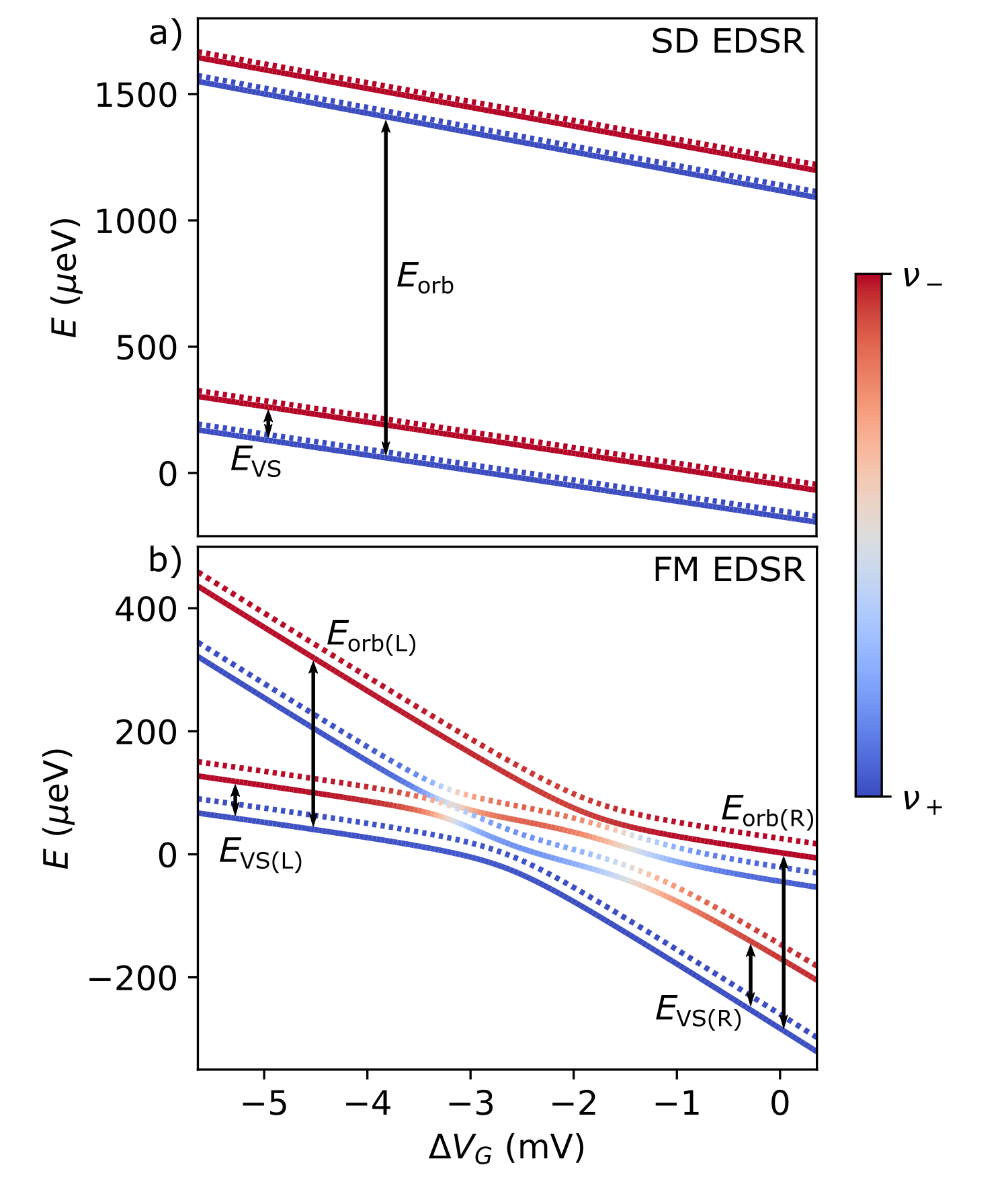}
		\caption{\textbf{Gate-voltage dependence of energy levels.} (a) Energy spectra for device \#2 in the SD EDSR regime plotted as a function of change in voltage from baseline on G$_{3}$. (b) Energy spectra for device \#2 in the FM EDSR regime plotted as a function of change in voltage from baseline on G$_{4}$. Solid (dashed) lines correspond to majority spin-up (spin-down) states, which are split in energy by $B_z \! \sim \! 23\,\mu$eV and the color of the lines denotes valley character.  Double-sided arrows indicate the orbital and valley splittings for the dot(s).} \label{fig:energy}
	\end{figure}

	The system can then be evolved according to the master equation
	\begin{flalign}
        \dot{\hat{\rho}}(t)&=-i\left[\hat{H}\left(\bv{V}(t),\bv{B}_z(t) \right),\hat{\rho}(t)\right]\nn\\
        &+\sum_i\mathcal{L}\left[\hat{\Gamma}_{V_i}\right]\hat{\rho}(t)+\mathcal{L}\left[\hat{\Gamma}_{\rm e-ph}\right]\hat{\rho}(t)\nn\\
        \hat{\Gamma}_{V_i}&=\sum_{n < m}\frac{\sqrt{4\pi S_{V_i}(\omega_{nm})}}{\hbar}\bra{n} \! \hat{H}_{V_i} \! \ket{m}\ket{n}\!\!\bra{m}\nn\\
        \hat{\Gamma}_{\rm e-ph}&= \sum_{n < m} \sqrt{\gamma_{nm}} \ket{n}\!\!\bra{m}\
	\end{flalign}
    with $\mathcal{L}\left[\hat{O}\right]\hat{\rho}=\hat{O}\hat{\rho}\hat{O}\dg-\frac{1}{2}\left\{ \hat{O}\dg\hat{O},\hat{\rho}\right\}$, where the electron-phonon-induced transition rates $\gamma_{nm}$ are given in Appendix \ref{appendix:Phonons} and
	where the voltage noise power spectra $S_{V_i}(\omega)$ are roughly determined by extending the $1/f$ spectrum of the low-frequency noise; we do not enforce continuity to reflect the fact that the high-frequency spectrum may significantly deviate from $1/f$ behavior, and, in particular, fall off faster than $1/f$. Specifically, at frequencies higher than about 1 GHz we choose $S_{V_i}(\omega)$ to be scaled by $0.5\times$ and $0.1\times$ the extended power spectral density in order to capture the impact of two possible $T_1$ regimes.

\section{Discussion and results}
	Performing numerical simulations for different voltage settings, we can construct an energy level diagram and generate the operators for the Hamiltonian and perturbations from noise.  The energy diagrams for a sample device are shown in Fig.~\ref{fig:energy}.  For the SD EDSR case, the lower two pairs of spin-split energy states correspond to different valley states in the ground orbital configuration, with the upper two pairs constituting the excited orbital states. For the FM EDSR case, to the far left or right of the avoided crossing a similar picture applies; \eg on the far left the lower two pairs of states belong to the left dot and are separated by the left dot valley splitting, and the upper states are on the right dot.  At the avoided crossing, the dot states hybridize with each other so valley-split sets of bonding/anti-bonding orbital pairs are obtained (each of which are spin split by the Zeeman field). We note the impact of alloy disorder on the electronic structure resulting in variations and asymmetry in the valley splitting and, in the case of the DQD regime, the offset of the avoided crossing.

	\begin{table}
		\begin{centering}
			\begin{tabular}{|c||c|c|c|}
				\hline
				\textbf{Device} &  \textbf{Valley} &  \textbf{Transition} & \textbf{Lever} \\
                                &  \textbf{Splitting} &  \textbf{Coupling} & \textbf{Arm} \\
				&  $E_{\mathrm{VS}}$  &  $\gamma_{V_{3}}$  & $\alpha_{v}$\\
                &  $\mu$eV &  $\mu$eV/mV &  $\mu$eV/mV\\
				\hline
				1 &  (71.6, 129.0) & (0.0067, 0.0032) & (3.3,1.3)\\
				2 &  (87.4, 303.6) & (0.0063, 0.0074) & (1.6,2.9)\\
				3 &  (170.2, 263.4) & (0.0027, 0.0027) & (0.6,0.5)\\
				4 &  (323.6, 254.6) & (0.0012, 0.0025) & (2.8,0.5)\\
				5 &  (253.5, 69.8) & (0.0034, 0.0028) & (1.7,5.3)\\
				\hline
			\end{tabular}
			\caption{\textbf{Single-dot EDSR operating parameters.} Key parameters for single quantum dots in five devices at the baseline electrode voltages with the quantum dot either under $G_2$ or $G_4$, respectively: the valley splitting which gives the energy gap between the two lowest spin-split pairs of energy states; the transition coupling $\gamma_{V_{3}} = \vert \langle E_{0} \vert \hat{H}_{{\rm G}_3} \vert E_{1} \rangle \vert$, which gives the coupling of the two lowest energy states due to changes in the voltage $V_{3}$ applied to drive electrode $G_{3}$; lever arm, which gives the change in energy gap due to changes in the drive electrode voltage.  }\label{tab:std_params}
		\end{centering}
	\end{table}

	\begin{table}
		\begin{centering}
			\begin{tabular}{|c||c|c|c|c|c|}
				\hline
				\textbf{Device} & \textbf{Tunnel}  & \textbf{Valley} & \textbf{Transition} & \textbf{Lever} \\
                                & \textbf{Coupling}  & \textbf{Splitting} & \textbf{Coupling} & \textbf{Arm} \\
                & $t_c$       & $(E_{\mathrm{VS}}^{\rm L},E_{\mathrm{VS}}^{\rm R})$ & $\gamma_{V_{4}}$& $\alpha_{v}$\\
                &  $\mu$eV        &  $\mu$eV & $\mu$eV/mV  &  $\mu$eV/mV\\
				\hline
				1 & 23.4 & (54.6, 116.3) & 3.1 & 83.1\\
				2 & 20.0 & (80.9, 238.5) & 3.2 & 75.1\\
				3 & 21.7 & (167.1, 262.0) & 3.5 & 84.9\\
				4 & 21.7 & (335.3, 253.8) & 3.2 & 81.9\\
				5 & 10.2 & (236.5, 33.3) & 24.4 & 80.8\\
				\hline
			\end{tabular}
			\caption{\textbf{Flopping-mode EDSR operating parameters.} Key parameters for double quantum dots in five devices at the baseline electrode voltages: the tunnel coupling, which depends on the energy barrier due to the central electrode and is given for the $n=1$ setting in Table \ref{tab:voltages}, the valley splitting for both the (left,right) dots, the transition coupling $\gamma_{V_{4}} = \vert \langle E_{0} \vert \hat{H}_{{\rm G}_4} \vert E_{1} \rangle \vert$, which gives the coupling of the two lowest energy states due to changes in the voltage $V_{4}$ of the drive electrode $G_{4}$, and lever arm, which gives the change in energy gap due to changes in the drive electrode voltage.}\label{tab:flp_params}
		\end{centering}
	\end{table}

	 	Selected Hamiltonian parameters for five device realizations are shown in Tables \ref{tab:std_params} and \ref{tab:flp_params} for SD and FM EDSR operation, respectively.  For the SD EDSR case, we give the valley splitting $E_{\mathrm{VS}}$, transition coupling $\gamma_{V_{3}} = \vert \langle E_{0} \vert \hat{H}_{{\rm G}_3} \vert E_{1} \rangle \vert$ between the two lowest states introduced by changes to the voltage on the drive gate $G_{3}$, and the valley lever arm $\alpha_{v}$ corresponding to the change in energy gap between the two lowest states with change in drive voltage. For the FM EDSR case, we calculate the tunnel coupling $t_c$, valley splitting for both dots $E_{\mathrm{VS}}^{\rm (L,R)}$, transition coupling $\gamma_{V_{4}} = \vert \langle E_{0} \vert \hat{H}_{{\rm G}_4} \vert E_{1} \rangle \vert$, and lever arm $\alpha_{v}$.  These parameters, especially the valley splitting, can vary significantly from device to device. These variations may have significant implications for FM EDSR operation, as the tunnel coupling that can be achieved is limited by the lowest of the two dots' valley splittings.  From Fig.~\ref{fig:energy}(b) we can see that as the energy gap between the two lowest orbital states $E_{\mathrm{orb}}^{\rm (L,R)}$ at the avoided crossing approaches either valley splitting, the excited valley state participates in hybridization and the avoided crossing becomes indistinct.  If either dot's valley splitting is low enough, the device cannot be operated in the flopping mode at all; this is the case for Device \#5 due to the low valley splitting of the right dot.
	
		Of particular interest are the lever arms, or how sensitive the energy gap between the lowest two states is to voltage fluctuations of this gate.  These are 0.5~--~5~$\mu$eV/mV for $G_3$ in SD EDSR mode, and around 80~$\mu$eV/mV for $G_4$ and 1--3~$\mu$eV/mV for $G_3$ in FM EDSR mode.  It is evident that, as expected, the FM of operation is significantly more sensitive to voltage fluctuations.

	\subsection{RB Performance}
		Both operating modes were evaluated by simulating single-qubit randomized benchmarking (RB)~\cite{Knill2008,Magesan2011,Magesan2012b}. 
        The single-qubit Cliffords were constructed by assembling sequences of primitive gates: $Z$ rotations -- furnished by idling -- and $X/Y$ rotations performed by pulsed ac driving of $V_3$ for SD EDSR and $V_4$ for FM EDSR~\cite{Magesan2012b}. 
        Pulses are flat with short (4 cycles) ramps up and down in amplitude.  These primitive gate sequences were optimized with respect to average fidelity in the absence of low-frequency noise and the remaining single-qubit Cliffords are assembled from these primitives.  In line with experimental protocols, RB simulations were performed by setting the initial state to the computational 0 state, applying a random sequence of single-qubit Cliffords terminated by the appropriate recovery gate, and projecting onto the computational 0 state \cite{Mills2022}.  
	
	We performed simulated RB experiments for circuit depths ranging from 1 to 10,000 gates.  For each circuit depth, we sampled 25 random Clifford circuits before moving on to the next depth. At the beginning and end of each circuit, we allotted 50~$\mu$s for state preparation and measurement (SPAM), during which the noise sources are allowed to evolve freely. For each device and mode (and for double-dot operation, setting of $V_3$), we chose drive amplitudes that result in a target Rabi frequency. We found that taking 25 random single-qubit Clifford gates per Clifford depth gives reasonable convergence of the inferred RB error rates.       
        
        \begin{figure}[ht]
		\centering
		\includegraphics[width=0.95\columnwidth]{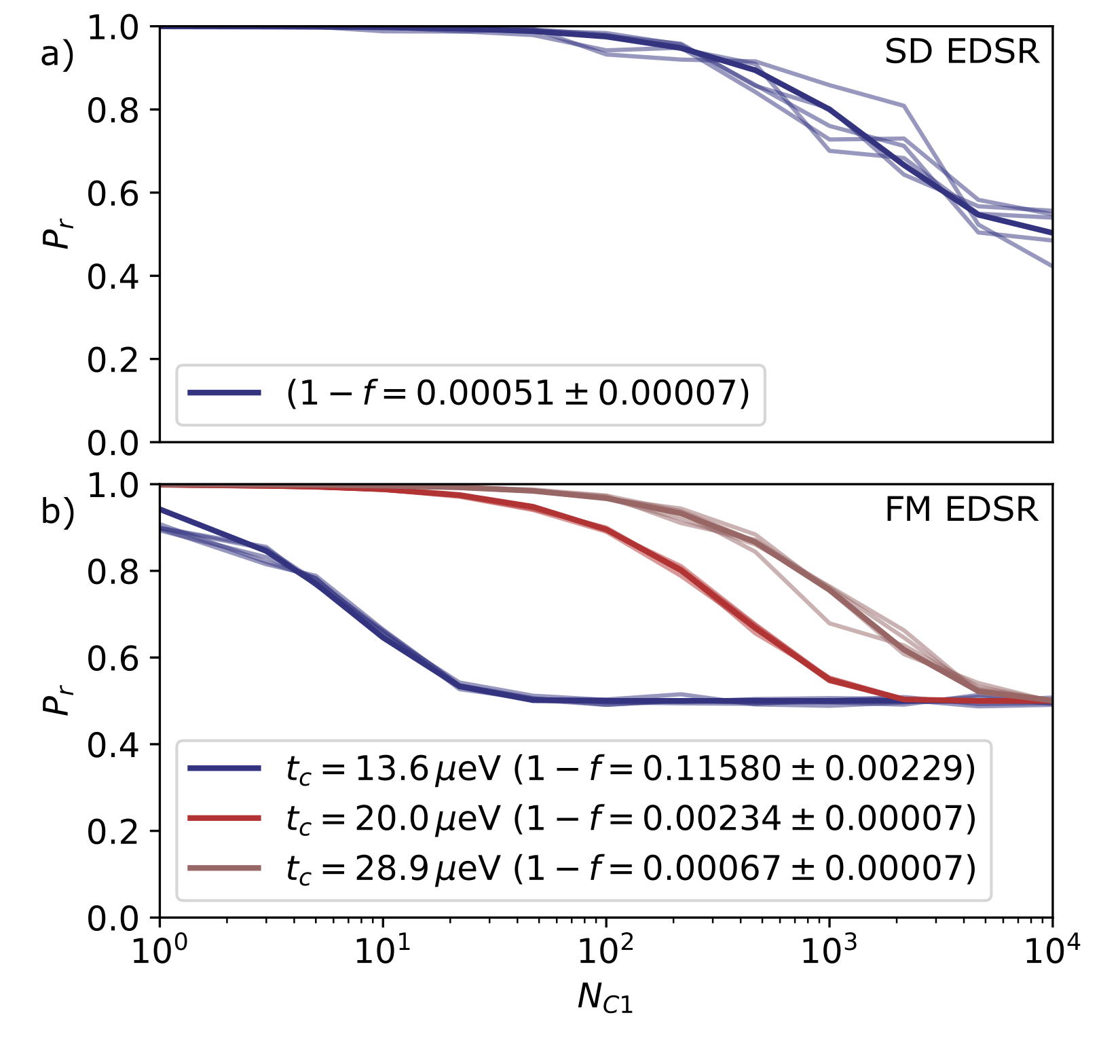}	
		\caption{ \textbf{Simulated randomized benchmarking return probabilities for SD and FM EDSR.} (a) Return probability $P_r$ as a function of the number of single qubit Clifford gates $C_{N1}$ for Device \#2 operated in SD EDSR mode. (b) Return probability $P_r$ as a function of the number of single qubit Clifford gates $C_{N1}$ for Device \#2 operated in FM EDSR mode. In both modes all sources of noise are included and the drive amplitudes $\Delta V$ were chosen such that the Rabi frequency is 3~MHz.  On each plot the  average infidelity is given, as well as $t_{c}$ for the FM EDSR case. The darker lines show the fitted average, while the lighter traces are the results for individual RB runs.} \label{fig:rb}
	\end{figure}
	Figure~\ref{fig:rb} shows our simulated RB results for Device \#2 under both modes of operation tuned to have a 3~MHz Rabi frequency.  The FM EDSR performance depends very strongly on the tunnel coupling -- and consequently the effective spin-orbit effect -- due to the increased charge noise at both high and low frequencies; higher tunnel coupling results in a more spin-like qubit with less noise sensitivity.  However, higher tunnel coupling also reduced the drive sensitivity, requiring larger drive amplitudes to achieve the targeted Rabi frequency.

\begin{figure*}[ht!]
	\centering
	\includegraphics[width=0.95\textwidth]{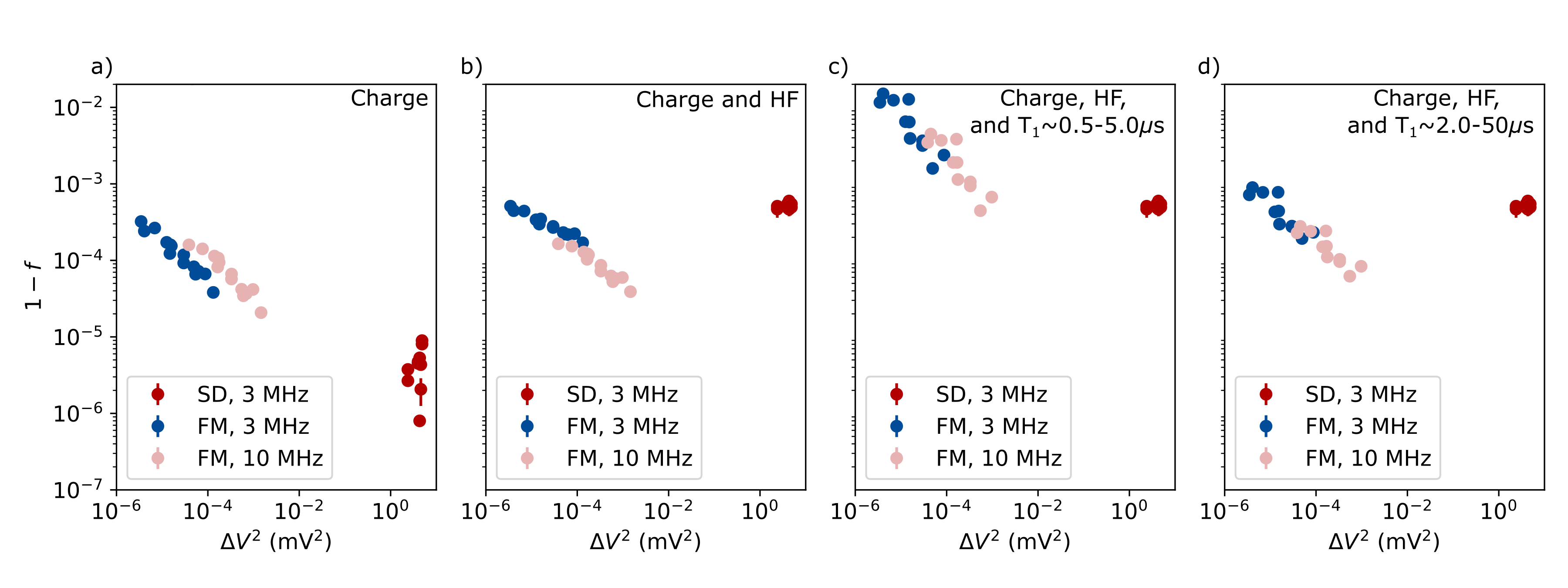}
	\caption{\textbf{Relation between infidelity and drive power.} Infidelity plotted as a function of drive amplitude squared, $\Delta V^2$, and aggregated across modeled device realizations for both SD and FM modes with various subsets of noise channels described in Sec \ref{sec:noise} included. For SD operation, device configurations with the dot under G$_2$ and G$_4$ are included, and for FM operation, multiple tunnel coupling strengths (obtained through different settings of $V_3$) for each device are included.  For both SD and FM, simulations are presented for operation calibrated to 3 MHz Rabi frequency (red and blue points, respectively); additionally, we include simulations for FM operation calibrated to a 10 MHz Rabi frequency (pink).} \label{fig:perfcomp}
\end{figure*}
    
	To understand this trade-off more broadly and trace the impact of different noise sources, in Fig.~\ref{fig:perfcomp} we plot infidelity versus drive amplitude squared (proportional to drive power) for all simulated device realizations and values of $t_c$ for both modes of operation at Rabi frequency 3~MHz, as well for additional tunings for the FM EDSR case with Rabi frequency set to 10~MHz, as noise channels are introduced.  We have omitted the FM EDSR RB for Device \#2 at $t_c=13.6$ due to the poor fit, as well as Device \#5 owing to a small valley splitting as described above. Generally, as expected, the infidelity improves with power.  Interestingly, in the presence of only charge noise this relationship is relatively consistent across devices and even the operating mode.  SD EDSR operation is largely insensitive to charge noise and enables the lowest infidelities, albeit at significantly higher power. With hyperfine noise added, the advantage of FM EDSR becomes clear. The performance in both regimes is ultimately hyperfine noise limited, with FM EDSR operation deriving an advantage from the wavefunction delocalization over both wells. Additionally, by operating in a higher Rabi frequency regime, the impact of hyperfine noise can be mitigated due to shorter gating times, allowing for infidelities an order of magnitude lower than the SD EDSR case, while still being operable at two orders of magnitude lower power. Unfortunately, $T_1$ noise proves to have a severe impact, and constitutes the dominant noise channel in FM EDSR mode. Achieving parity in infidelity with SD EDSR requires $T_1 \sim30~\mu$s at a 3~MHz Rabi frequency, and $T_1 \sim~5~\mu$s at a 10~MHz Rabi frequency; however, this occurs at less than 1/1000th of the drive power required for SD EDSR.

\section{Conclusion}
\label{sec:Conclusion}
Starting from detailed device models, including Si/SiGe quantum wells generated by explicit modeling of germanium defects and multiple relevant noise channels, we have simulated randomized benchmarking of single-dot ``standard'' and double-dot ``flopping-mode'' operation with temporally correlated noise. We analyzed the impact of different noise sources on the performance of both operating modes, allowing for comparative evaluation.  We find that while the increased sensitivity of FM EDSR operation does allow for significantly lower power operation, it comes at the cost of increased charge noise exposure.  $T_1$ noise in particular has a dominant impact on performance; devices must have sufficiently long $T_1$ times to enable fidelity comparable to that of single-dot operation, though higher driving frequencies can mitigate this.  Importantly, the required $T_1$ times are near the range of experimentally observed values.
Additionally, by investigating multiple devices we gain insight into the impact of variation in alloy disorder from device to device.  For single-dot operation, the impact is minimal, however, double-dot performance can vary widely.  Variation in valley-splitting is especially large and places a ceiling on the attainable tunnel couplings, which can constrain a given device's performance potential and even prohibit FM EDSR operation altogether. In the future, we expect that it would be informative to investigate in more detail the influence of alloy disorder on device yield.

\acknowledgments
This article has been co-authored by an employee of National Technology \& Engineering Solutions of Sandia, LLC under Contract No. DE-NA0003525 with the U.S. Department of Energy (DOE). The employee owns all right, title and interest in and to the article and is solely responsible for its contents. The United States Government retains and the publisher, by accepting the article for publication, acknowledges that the United States Government retains a non-exclusive, paid-up, irrevocable, world-wide license to publish or reproduce the published form of this article or allow others to do so, for United States Government purposes. The DOE will provide public access to these results of federally sponsored research in accordance with the DOE Public Access Plan \url{https://www.energy.gov/downloads/doe-public-access-plan.} This work was performed, in part, at the Center for Integrated Nanotechnologies, an Office of Science User Facility operated for the U.S. Department of Energy (DOE) Office of Science. JRP acknowledges the support of ARO Grant W911NF-23-1-0104 and AFOSR Grant FA9550-23-1-0710.

\appendix
\section{Envelope function approximation}
\label{appendix:Laconic}
We model the electronic structure of our quantum dot including valley physics for explicit alloy disorder using an envelope function approximation (EFA) as our starting point.
The EFA is defined by a set of multi-valley effective mass equations~\cite{Pena2024}, and these equations are handled numerically using discontinuous Galerkin (DG) methods~\cite{HesthavenWarburtonBook}.
In particular, we use a polynomial basis for an interior penalty discretization of the Hamiltonian to build a matrix that can be diagonalized for wavefunctions and their associated energies.

To capture the effects of valley-orbit coupling with atomistic details, we combine our envelope functions with their associated Bloch functions in order to bridge the atomic scale details with the features of the full quantum dot wavefunction.
The Ge concentration profile is used to generate the locations of Ge atoms, the effects of which are then approximated by repulsive delta functions.
For $u_\mu(\mathbf{r})$ as the Bloch function for the valley at $\mathbf{k}_\mu$ in momentum space, $F_\mu$ as the envelope function for valley $\mu$, and $V_{\textrm{tot}}$ being the total potential from electrostatic gates as well as the delta functions at the Ge sites, the total valley-orbit coupling is determined by
\begin{multline}
    \bra{\psi_\mu}\hat{V}_{\textrm{tot}}\ket{\psi_\nu}=\\\int\!d^3\mathbf{r}\,e^{i\left(\mathbf{k}_\mu-\mathbf{k}_\nu\right)\cdot\mathbf{r}}u_\mu^*(\mathbf{r})u_\nu(\mathbf{r})F_\mu^*(\mathbf{r})V_{\textrm{tot}}(\mathbf{r})F_\nu(\mathbf{r}),
\end{multline}
where $\psi_\mu=u_\mu F_\mu$.
This equation is used to incorporate the full valley-orbit coupling into the DG discretization of the potential.
Further details about the delta functions representing the effects of Ge atoms, Bloch functions, and externally applied potential follow those found in Ref.~\cite{Pena2024}.

\section{Modeling electron-phonon coupling}
\label{appendix:Phonons}
Given a solution for the energy eigenstates of an electron in the heterostructure at a given operating voltage, we now describe how we model the effects of electron-phonon coupling. In this work, we make the approximation that relevant acoustic phonon modes in the Si well are those of bulk Si with the further simplification of having isotropic dispersion \cite{Pop2004}. While the SiGe alloy, Si/SiGe interfaces, and other material boundaries such as between SiGe/dielectric and dielectric/electrode may add further structure to the phonon spectrum, we take this level of complexity to be beyond the scope of the present work. Given that relevant energy scales here are $\mathcal{O}(\mathrm{meV})$ or lower, the phonon modes participating in relaxation are the longitudinal and transverse acoustic modes.

To account for the influence of phonons, we consider the electron spin and lattice as an open quantum system given by a spin-boson model \cite{Breuer2002}. The electron-phonon interaction takes the form \cite{Hasegawa1960,Raith2012}
\begin{equation}
    H_{\mathrm{e-ph}}= \!\!\!\!\sum_{\mathbf{q},\lambda\in\{\mathrm{L},\mathrm{T}\}} \!\!\!\! \xi^{\lambda}(\mathbf{q})e^{i\mathbf{q}\cdot\mathbf{r}}\hat{b}_{\lambda,\mathbf{q}}^{\dagger}+\xi^{\lambda*}(\mathbf{q})e^{-i\mathbf{q}\cdot\mathbf{r}}\hat{b}_{\lambda,\mathbf{q}}
\end{equation}
where for an electron in the $\pm z$ valley states we have
\begin{eqnarray}
    \xi^{L}(\mathbf{q}) & = & i\sqrt{\frac{\hbar q}{2\rho Vc_{L}}} \left( \Xi_{d}+\Xi_{u}\cos^{2}(\theta) \right) \\
    \xi^{T}(\mathbf{q}) & = & i\sqrt{\frac{\hbar q}{2\rho Vc_{T}}} \Xi_{u}\sin(\theta)\cos(\theta)
\end{eqnarray}
with $\Xi_{d} = 1.1 \ \mathrm{eV}$ the dilatation and $\Xi_{u} = 10.5 \ \mathrm{eV}$ the shear deformation potentials \cite{Fischetti1996}. We assume a density for Si of $\rho = 2329 \ \mathrm{kg/m^{3}}$. We approximate the phonon dispersion for mode $\lambda$ at wavevector $q = \Vert \mathbf{q} \Vert$ as isotropic, $\omega_{\lambda,q} = \omega_{\lambda,0} + c_{\lambda} q + d_{\lambda} q^{2}$, with parameters given by Ref \cite{Pop2004}.

Making standard approximations to derive a Markovian master equation \cite{Breuer2002}, we find the following transition rate from eigenstates $\vert E_{m} \rangle$ to $\vert E_{n} \rangle$ having energy difference $E_{m} - E_{n} = \hbar \omega_{mn}$,
\begin{equation}
    \gamma_{nm} =\frac{V}{\hbar^{2}(2\pi)^{2}}\sum_{\lambda}\int d^{3}\mathbf{q}\ \vert g_{nm}^{\lambda}(\mathbf{q})\vert^{2}\delta(\omega_{mn}-\omega_{\lambda,\mathbf{q}}), \label{eq:BareRelaxRate}
\end{equation}
where $\lambda \in \lbrace \mathrm{L}, \mathrm{T} \rbrace$ indexes the longitudinal and transverse acoustic modes and the matrix element
\begin{equation}
    g_{nm}^{\lambda}(\mathbf{q})=\langle E_{n}\vert \xi^{\lambda}(\mathbf{q})e^{i\mathbf{q}\cdot\mathbf{r}}\vert E_{m}\rangle
\end{equation}
quantifies the mode- and wavevector-dependent coupling between eigenstates. 
Having assumed isotropic phonon dispersion, the integral in Eq \ref{eq:BareRelaxRate} simplifies to a 2D on-shell integral over the angular coordinates $\theta,\varphi$. We evaluate this 2D integral numerically using 10$^\textrm{th}$-order Gaussian quadrature along each axis. Note that spin-orbit coupling is included by virtue of the eigenstates $\vert E_{n} \rangle$ having both valley-orbital and spin character.

\section{Modeling charge noise-induced $T_1$ processes}
\label{appendix:ChargeNoiseT1}
In addition to electron-phonon coupling, it is known that charge noise also drives spin $T_1$ processes \cite{Huang2014,Borjans2019,Hosseinkhani2022}. The spin relaxation rate is given by 
$1/T_{1} = \sum_{i} \frac{8 \pi^{2}}{\hbar^2} S_{V_{i}}(f) \vert \langle E_{f} \vert \hat{H}_{G_{j}} \vert E_{i} \rangle \vert^{2}$, where $j$ indexes each gate electrode, $V_{j} \hat{H}_{G_{j}}$ is the Hamiltonian contribution to the electrostatic potential from each gate, and $\vert E_{i} \rangle$, $\vert E_{f} \rangle$ are the relevant qubit spin-valley-orbital eigenstates.

\bibliography{refs}

\end{document}